\documentclass[acmsmall]{acmart}

\usepackage{subcaption}
\usepackage{amsmath}
\usepackage{hyperref}
\usepackage{cleveref}
\usepackage{alltt}
\usepackage{tikz}
\usepackage{float}
\floatstyle{ruled}
\restylefloat{figure}

\newcommand{\fsm}{\texttt{FSM}}
\newcommand{\dfa}{\texttt{dfa}}
\newcommand{\ndfa}{\texttt{ndfa}}

\newcommand{\sig}{\texttt{\(\Sigma\)}}

\newcommand{\delt}{\texttt{\(\delta\)}}
\newcommand{\lamb}{\texttt{$\lambda$}}
\newcommand{\quot}{\texttt{\textquotesingle{}}}

\newcommand{\steps}{\texttt{$\vdash^*$}}
\newcommand{\elist}{\texttt{\textquotesingle{()}}}
\newcommand{\qquot}{\texttt{\textasciigrave{}}}

\newcommand{\arrow}{\(\rightarrow\)}
\newcommand{\flatt}{\texttt{FLAT}}
\newcommand{\ets}{\texttt{$\epsilon$}-transitions}
\newcommand{\et}{\texttt{$\epsilon$}-transition}
\newcommand{\vdotss}{\(\vdots\)}
\newcommand{\ep}{\texttt{\(\epsilon\)}}
\newcommand{\gviz}{\texttt{GraphViz}}

\AtBeginDocument{%
  }

\begin{document}

\title{Visualizing a Nondeterministic to Deterministic Finite-State Machine Transformation}


\author{Tijana Minic}
\affiliation{\institution{Seton Hall University}
             \city{South Orange}
             \state{NJ}
             \country{USA}
             \postcode{07079}}
\email{minictij@shu.edu}

\author{Marco T. Moraz\'{a}n}
\affiliation{%
  \institution{Seton Hall University}
  \city{South Orange}
  \state{NJ}
  \country{USA}}
\email{morazanm@shu.edu}

\renewcommand{\shortauthors}{Moraz\'{a}n and Minic}

\begin{abstract}
The transformation of a nondeterministic finite-state automaton into a deterministic finite-state automaton is an integral part of any course on formal languages and automata theory. For some students, understanding this transformation is challenging. Common problems encountered include not comprehending how the states of the deterministic finite-state automaton are determined and not comprehending the role that all the edges of the nondeterministic finite-state automaton have in the deterministic finite-state automaton's construction. To aid students in understanding, transformation visualization tools have been developed. Although useful in helping students, these tools do not properly illustrate the relationship between the states of the deterministic finite-state automaton and the edges of the nondeterministic finite-state automaton. This article presents a novel interactive visualization tool to illustrate the transformation that highlights this relationship and that is integrated into the \fsm{} programming language. In addition, the implementation of the visualization is sketched.
\end{abstract}






\maketitle

\section{Introduction}

Formal Languages and Automata Theory (\flatt{}) courses and textbooks place a great deal on emphasis on constructive proofs. Constructive proofs are commonly used to establish, for example, the equivalence of different computation models. As part of such proofs, a construction algorithm is designed that transforms one model of computation into another. \flatt{} students are commonly exposed to the transformation of a nondeterministic finite-state automaton (\ndfa) into a deterministic finite-state automaton (\dfa). It is important for students to learn about such a transformation, because \ndfa{}s are usually easier to design than \dfa{}s and \dfa{}s are usually easier to implement using a programming language. Therefore, learning about the \ndfa{} to \dfa{} transformation makes it easier to design and write software for string searching \cite{baase} and spell checking \cite{spellchecks} in word processors, for recognizing tokens, keywords, and identifiers in compilers \cite{parsons,lexanalysis}, for specifying communication event sequences in network design \cite{prithi}, for detecting malicious behavior by cybersecurity software \cite{ilgun}, and for formal software verification \cite{gheorghe}.

More often than desirable, students struggle to understand the \ndfa{} to \dfa{} transformation. This is not entirely surprising, because a path in a \dfa{} transition diagram may represent multiple paths in an \ndfa{} transition diagram. Thus, a state in the \dfa{} under construction may represent any subset of states in the \ndfa{} and an edge in the \dfa{} under construction may represent multiple \ndfa{} edges. We shall call the states in the \dfa{} under construction \emph{super states}. Mentally or manually tracking super states and path information during the transformation process is taxing and error-prone. To aid students, visualization tools have been developed. These tools build the \dfa{} transition diagram by adding edges in a stepwise manner. Rarely, however, is proper attention given to where super states come from and to the \ndfa{} edges that are represented by a \dfa{} edge. Therefore, there is a need to improve the visualization of this transformation.

This article describes a new visualization tool for the \ndfa{} to \dfa{} transformation developed for \fsm{}\footnote{\fsm{} is available for free on GitHub: \href{https://github.com/morazanm/fsm} {\fsm{}} .}--a domain-specific language for the \flatt{} classroom \cite{fsm,fsm-viz}. With this new tool, the user may visualize the stepwise addition of \dfa{}-edges while simultaneously visualizing its corresponding \ndfa{}-edges. In addition, the tool also highlights the \ndfa{} edges that are already captured by the \dfa{} under construction. Users may step forward and backward in the \dfa{} construction as well as jump directly to the end of the \dfa{} construction to view the result. The article is structured as follows. \Cref{rw} briefly reviews and contrasts with related work. \Cref{fsm-description} introduces \fsm{} and the constructors for \dfa{}s and \ndfa{}s. \Cref{fsm-transformation} describes the implementation of the transformation algorithm in \fsm{}. \Cref{fsm-viz} describes the transformation's visualization. Finally, \Cref{concl} presents concluding remarks and directions for future work.

\section{Related Work}
\label{rw}

\subsection{The Transformation}

\flatt{} textbooks usually hinge the conversion from an \ndfa{}, \texttt{N=(K \sig{} s F \delt{})}, to a \dfa{}, \texttt{D=(K$^{\prime}$ \sig{}$^{\prime}$ s$^{\prime}$ F$^{\prime}$ \delt{}$^{\prime}$)}, on the observation that \texttt{N} may occupy multiple states during a computation \cite{gurari,hopcroft,lewis,martin,rich,sipser}. Namely, the states that \texttt{N} may occupy are the states reachable with the consumed input. To understand which are the reachable states given some input, it is useful to introduce \texttt{E(q)}, the notion of the empties of \texttt{q}, where \texttt{q}$\in$\texttt{K}. Informally, the empties of \texttt{q} is the set of states reachable from \texttt{q} using only empty transitions. More formally, the empties of \texttt{q} is defined as follows:
\begin{alltt}
     E(q) = \{p | ((\(\epsilon\) q) \steps (\(\epsilon\) p))\}
\end{alltt}

\texttt{D} is constructed using the following algorithm:
\begin{description}
  \item[K$^{\prime}$ = ] 2$^K$, the power set of K
  \item[\sig{}$^{\prime}$ =] \sig{}
  \item[s$^{\prime}$ =] E(s)
  \item[F$^{\prime}$ =] \{B $|$ B$\subseteq$K$^{\prime}$ $\wedge$ B$\cap$F$\neq \emptyset$\}
  \item[\delt{}$^{\prime}$ =] \{(Q y $\bigcup$E(p)) $|$ p$\in$K $\wedge$ (q y p)$\in$\delt{} $\wedge$ Q$\in$K$^{\prime} \wedge$ q$\in$Q\}
\end{description}
The states in \texttt{K}$^{\prime}$ are, as mentioned, usually refered to as super states given that each may contain more than one state in \texttt{K}. The algorithm is correct, but it is also terse and, given that the power set of a finite set needs to be computed, not practical for other than small sets of states. Most textbooks, therefore, observe that not all the super states in K$^{\prime}$ are relevant and that the only relevant super states are those that are reachable from \texttt{E(s)}. Under this mantle, \dfa{} construction becomes a search for a transition function on super states starting from \texttt{E(s)}.

In contrast, the \fsm{} transformation from \ndfa{} to \dfa{} is described and implemented from the start as a search for a transition function on super states. In this manner, students do not need to worry about computing or searching the power set of a set. Instead, they may focus on a breadth-first-based algorithm that searches for the needed super-state transitions. This algorithm requires the computation of a table for the empties of every state in the given \ndfa{} and the computation of an encoding table that maps super states to valid state names in \fsm{}. Such a programming approach appeals to Computer Science students that wish to hone their programming skills. This algorithm is further described in \Cref{fsm-transformation}.

\subsection{Visualization}

\begin{figure}[t!]
\centering
\begin{subfigure}{0.49\textwidth}
\centering
\includegraphics[scale=0.2]{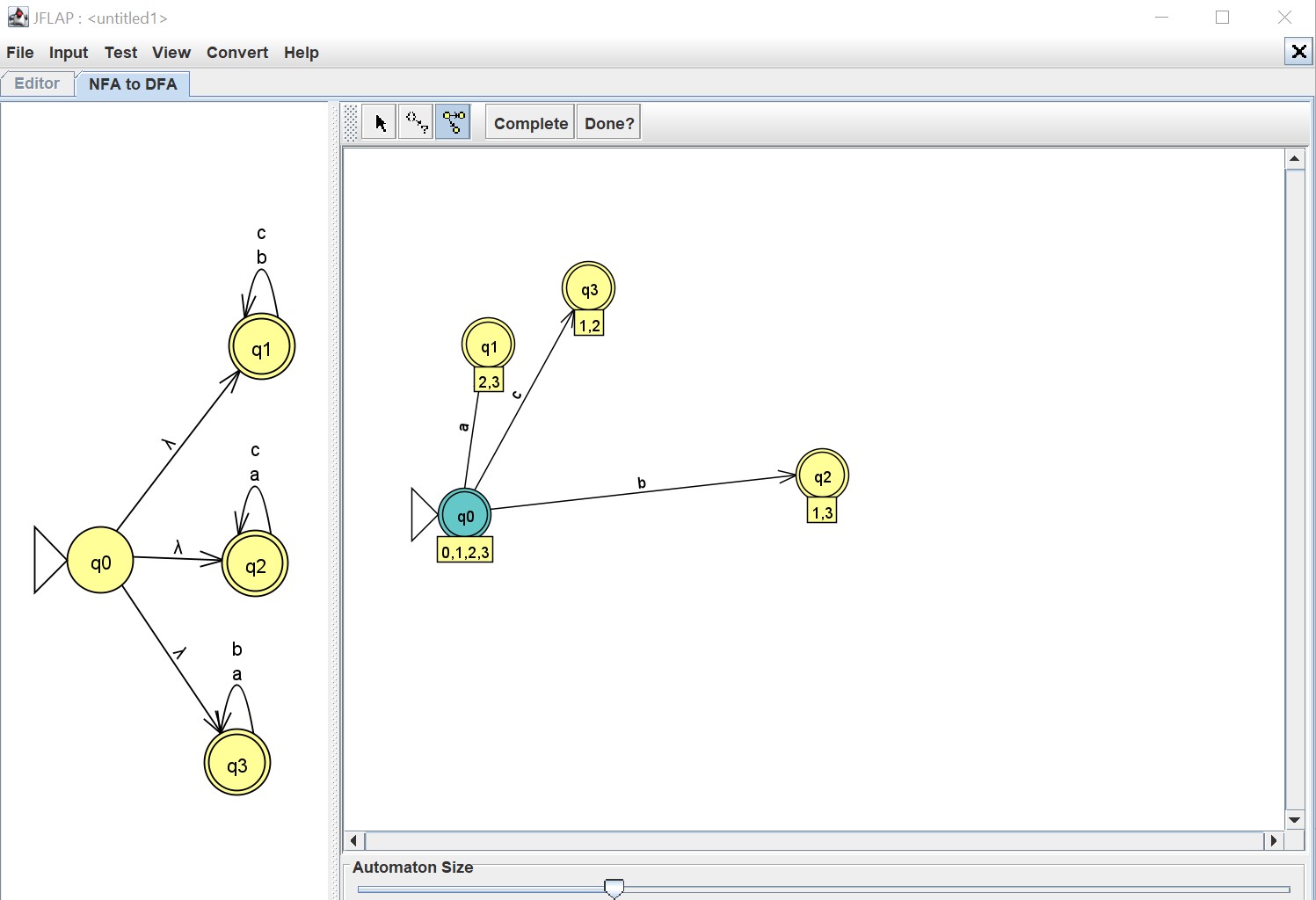}
\caption{After expanding the starting state.}
\label{jflap1}
\end{subfigure}
\hfill
\begin{subfigure}{0.49\textwidth}
\centering
\includegraphics[scale=0.2]{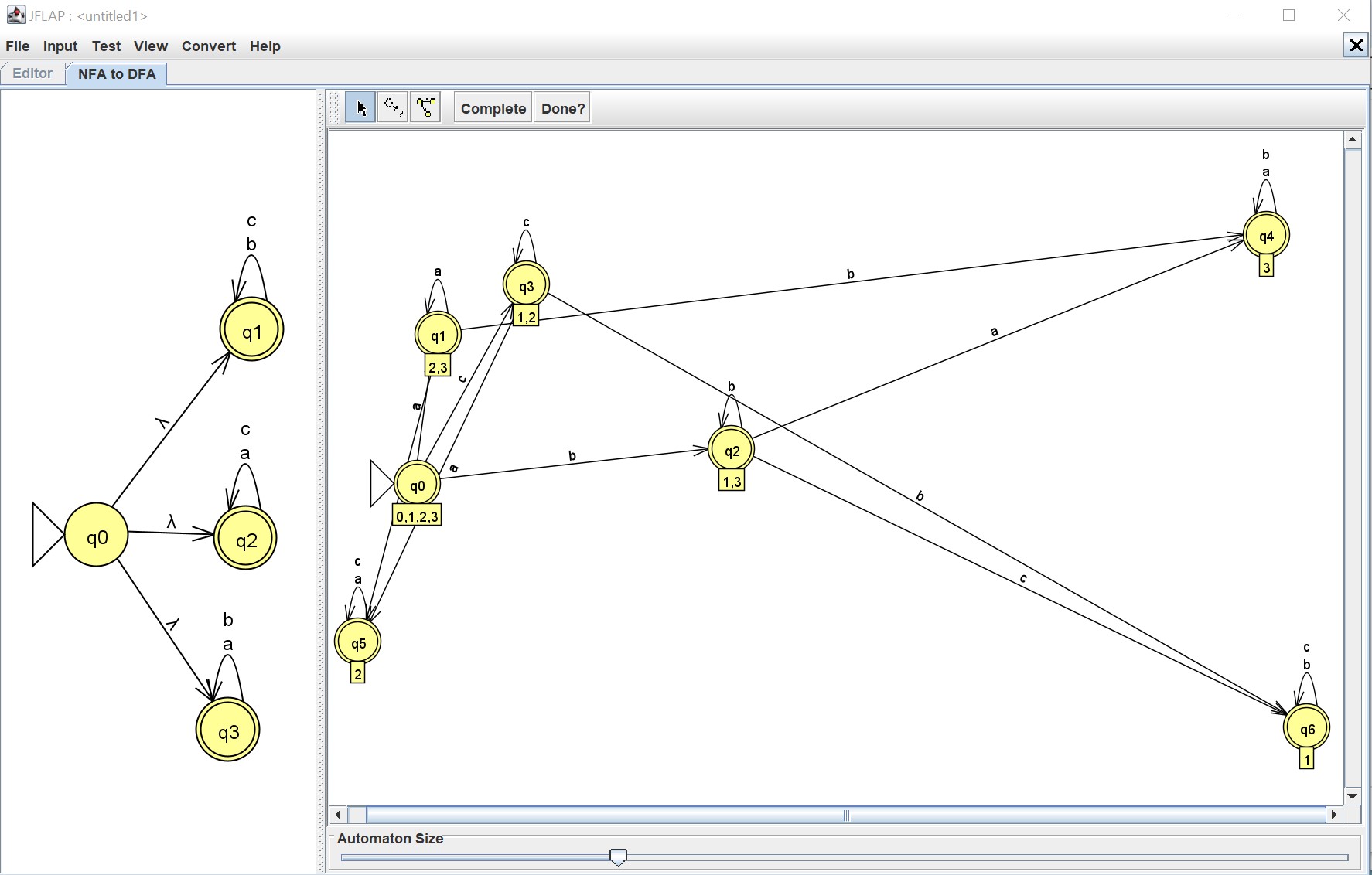}
\caption{Completed Transformation.}
\label{jflap2}
\end{subfigure}
\caption{\texttt{JFLAP} visualization.}
\end{figure}

\texttt{JFLAP} is a visualization tool that allows users to graphically edit \ndfa{}s and convert them to \dfa{}s \cite{rodger,rodgerII}. The conversion may be done piecemeal on node at a time or may be done in one step. Piecemeal conversion, for example, allows users to click on a super state to generate all the edges out of the clicked super state. \Cref{jflap1} displays the first such step in such a conversion. The \ndfa{} on the left decides the language \texttt{(b $\cup$ c)$^*$ $\cup$ (a $\cup$ c)$^*$ $\cup$ (a $\cup$ b)$^*$}. The transition diagram for the \dfa{} starts only containing the initial super state. After the initial super state is expanded, three edges are added. Completing the rest of the transformation in one step yields the configuration displayed in \Cref{jflap2}. There are salient features that merit scrutiny. The first is that the \dfa{} transition diagram in \Cref{jflap2} does not represent a function. It is missing transitions from \texttt{q4} on an \texttt{a}, from \texttt{q5} on a \texttt{b}, and from \texttt{q6} on a \texttt{c}. The missing transitions, clearly, all ought to go to a dead state. This brings us to the second feature that merits scrutiny: there is a missing state. Namely, the dead state. These features can be problematic with a student just starting to study \flatt{}. After studying the transformation algorithm, they naturally question why every state does not have a transition for every letter in the alphabet. The third feature that merits scrutiny is the layout of the \dfa{} transition diagram. It is unappealing and difficult to read. The fourth issue is that it is not clear, when the graph is constructed piecemeal, which \ndfa{} edges have been used to construct the \dfa{} and which are yet to be used. Finally, the fifth issue is that the piecemeal transformation only runs in one direction. The user is unable to step back to review previous steps.

\begin{figure}[t!]
\centering
\begin{subfigure}{0.49\textwidth}
\centering
\includegraphics[scale=0.30]{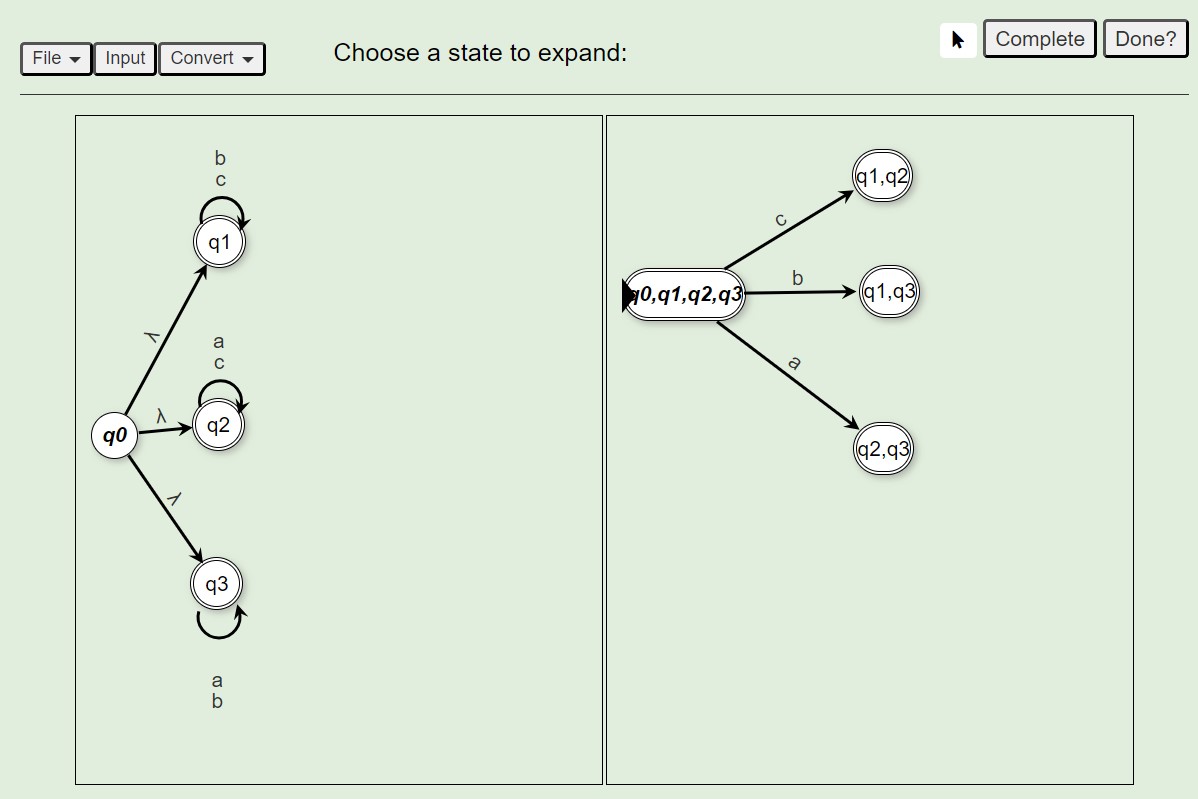}
\caption{After expanding the starting state.}
\label{openflap1}
\end{subfigure}
\hfill
\begin{subfigure}{0.49\textwidth}
\centering
\includegraphics[scale=0.30]{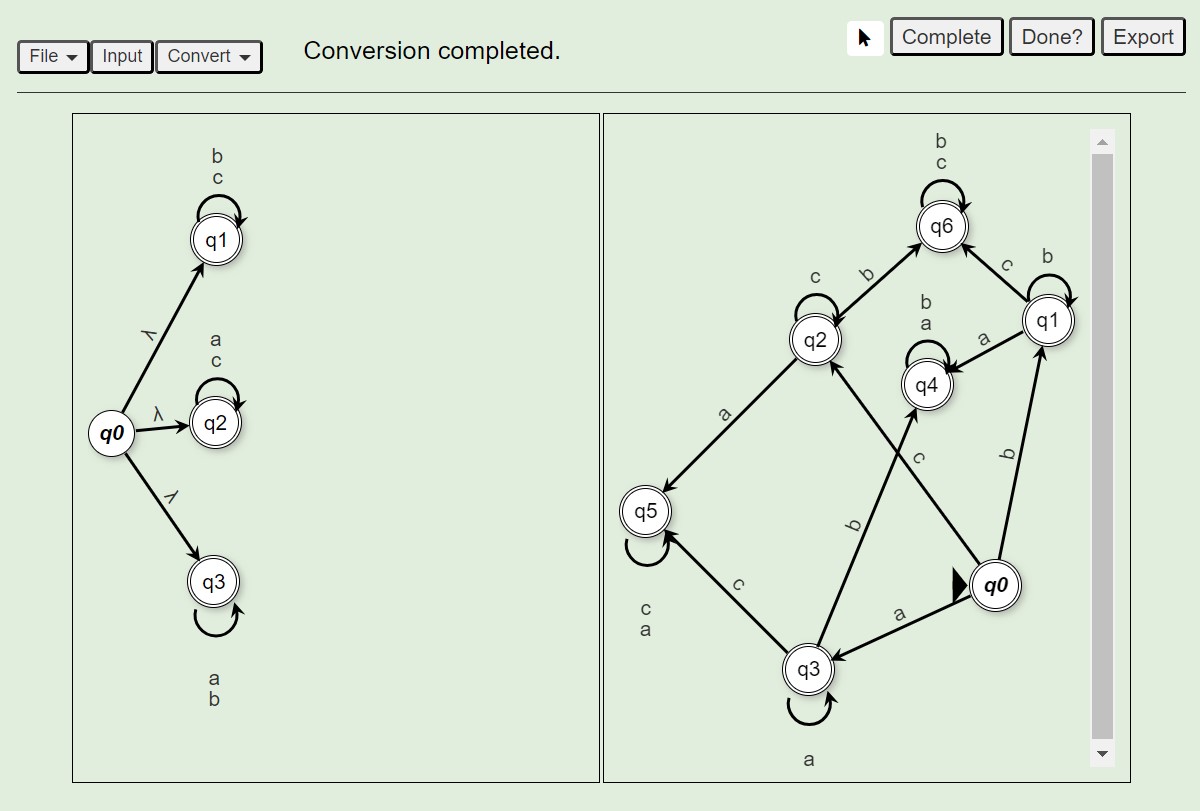}
\caption{Completed Transformation.}
\label{openflap2}
\end{subfigure}
\caption{\texttt{OpenFLAP} visualization.}
\end{figure}

\texttt{OpenFLAP} is another visualization tool that allows users to graphically edit \ndfa{}s and convert them to \dfa{}s \cite{mohammed}. This tool demands more interaction with the user when the \dfa{} is built piecemeal. As in \texttt{JFLAP}, the state diagram for the \dfa{} starts with \texttt{E(s)}. When a user clicks on a state, she is prompted for the alphabet element to expand on. The user then may place a super state, but first must specify the super state's member \ndfa{} states. The first three steps of the piecemeal conversion are displayed in \Cref{openflap1}. Observe that the layout is nicer and easier to read than what is produced by \texttt{JFLAP} as displayed in \Cref{jflap1}. The layout is better because the user has chosen to carefully place the super states. Like \texttt{JFLAP}, \texttt{OpenFLAP} allows for the \dfa{}'s completion in one step. The result of such a step is displayed in \Cref{openflap2}. As with \texttt{JFLAP}, the layout is unappealing and difficult to read, the dead state and associated transitions are missing, and, when the \dfa{} is built piecemeal, it is not easy to determine which \ndfa{} edges have been used to construct the \dfa{} and which are yet to be used. Finally, the user is also unable to step back in the piecemeal computation to review previous steps.

In contrast, the work presented in this article completely automates the piecemeal construction of the \dfa{}'s state transition diagram. The user cannot choose the direction of the expansion. Instead, she can step backwards and forwards in the computation. In addition, the user may complete the transformation in one step and still step backwards. Furthermore, \gviz{} \cite{graphviz} is used to layout the \ndfa{} and the \dfa{} transition diagrams in an appealing manner. When an edge is added to the \dfa{}, the \ndfa{} edges it accounts for are highlighted in the \ndfa{}'s transition diagram. In this manner, students can easily determine how \ndfa{} and \dfa{} edges correspond. Any \ndfa{} edges previously used in the \dfa{}'s construction are faded, and \ndfa{} edges not yet used in the construction are rendered as solid black edges. Finally, in further contrast with \texttt{JFLAP} and \texttt{OpenFLAP}, the \dfa{} transition graph is complete. That is, it includes a dead state, and associated transitions when needed. The \fsm{} transformation visualization is further discussed in \Cref{fsm-viz}.

\section{Brief Introduction to \fsm{}}
\label{fsm-description}

\subsection{Core Definitions}

In \fsm{}, a state is an uppercase Roman alphabet symbol, and an input alphabet symbol is a lowercase Roman alphabet letter or number. An input alphabet, \texttt{$\Sigma$}, is represented as a list of alphabet symbols. A word is either \texttt{EMP}, denoting the empty word, or a nonempty list of alphabet symbols. Each word is given as input to a state machine to decide if it is in the machine's language.

The machine constructors of interest for this article are those for finite-state automatons:
\begin{alltt}
     make-dfa: K \sig{} s F \delt{} [\quot{}no-dead] \arrow{} dfa
     make-ndfa: K \sig{} s F \delt{} \arrow{} ndfa
\end{alltt}
Here, \texttt{K} is a list of states, \texttt{F} is a list of final states in \texttt{K}, \texttt{s} is the starting state in \texttt{K}, and \texttt{$\delta$} is a transition relation. A transition relation is represented as a list of transition rules. This relation must be a function for a \dfa{}. A \dfa{} transition rule is a triple, \texttt{(K $\Sigma$ K)}, containing a source state, the element to read, and a destination state. The optional \texttt{\quot{}no-dead} argument for \texttt{make-dfa} indicates to the constructor that the relation given is a function. Omitting this argument indicates that the transition function is incomplete and the constructor adds a fresh dead state along with transitions to this dead state for any missing transitions. An \ndfa{} transition rule is a triple, \texttt{(K $\{\Sigma \cup \{EMP\}\}$ K)}, containing a source state, the element to read (possibly none), and a destination state.

Observers are functions that use a given machine to compute a result. The observers are:
\begin{alltt}
     (sm-states M)  (sm-sigma M) (sm-start M) (sm-finals M)  (sm-rules M)
     (sm-type M)
     (sm-apply M w)
     (sm-showtransitions M w)
\end{alltt}
The first 5 observers extract a component from the given state machine. The sixth returns the given state machine's type. The seventh applies the given machine to the given word and returns \texttt{\quot{}accept} or \texttt{\quot{}reject}. The eighth returns a trace of the configurations traversed applying the given machine to the given word ending with the result. A trace is only returned, however, if the machine is a \dfa{} or if the word is accepted by an \ndfa{}.

\subsection{Visualization}

\fsm{} provides machine rendering and machine execution visualization. The current visualization primitives are:
\begin{alltt}
     \texttt{(sm-graph M)}     \texttt{(sm-visualize M [(s p)\(\sp{*}\)])}
\end{alltt}
The \texttt{(sm-graph M)}    returns an image of the given machine's transition diagram. In the returned image, a node represents a state and an edge represents a transition rule. Each node is a state enclosed in a circle variety. The starting state is denoted by a green circle. A final state is denoted by a double black circle. If the starting state is also a final state then it is denoted using a double green circle. All other states are denoted using a single black circle. The label on an edge denotes the element that is consumed using the rule the edge represents.

\begin{figure}[t!]
\centering
\begin{subfigure}[b]{0.49\textwidth}
\centering
\includegraphics[scale=0.3]{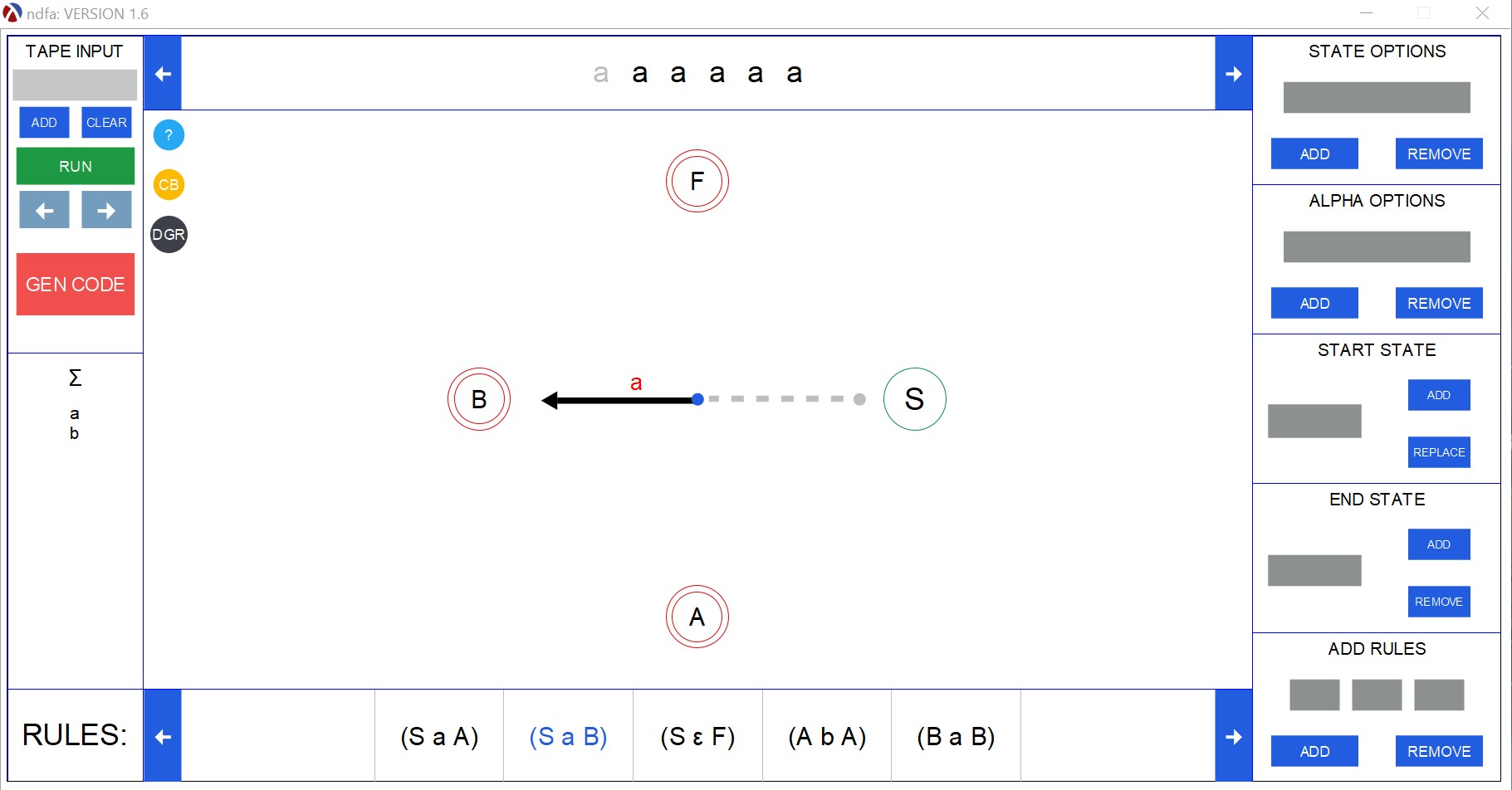}
\caption{Control view.} \label{viztool1}
\end{subfigure}
\hfill\\
\begin{subfigure}[b]{0.49\textwidth}
\centering
\includegraphics[scale=0.3]{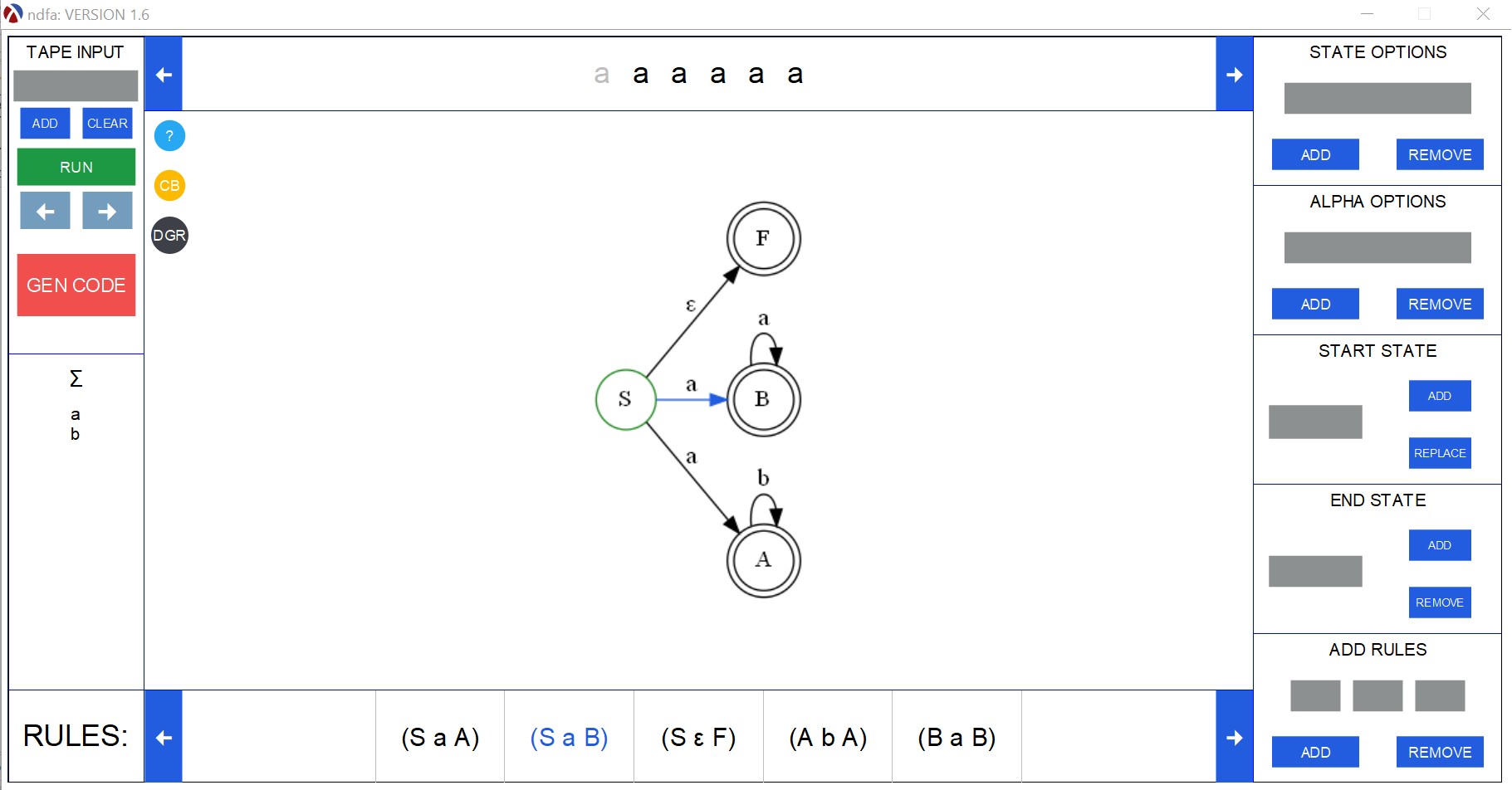}
\caption{State diagram view.} \label{viztool2}
\end{subfigure}
\caption{Execution Visualization for an \ndfa{} Deciding L=\{$\epsilon \cup aa^{\texttt{*}} \cup ab^{\texttt{*}}$\}} \label{e-aa-start-ab-star}
\end{figure}


The \texttt{(sm-visualize M [(s p)\(\sp{*}\)])} launches the \fsm{} visualization tool to simulate machine execution. The optional two-element lists contain a state of the given machine and an invariant predicate for the state. Machine execution may always be visualized if the machine is a \dfa{}. Similarly to \texttt{sm-showtransitons}, \ndfa{} machine execution may only be visualized if the given word is in the machine's language. Within the visualization tool, a machine is depicted in control view or in state diagram view. In control view, as depicted in \Cref{viztool1}, final states are denoted using double red circles. The arrow points to the current state and the dashed line indicates the previous state. In \Cref{viztool1}, the last rule used is \texttt{(S a B)}. In state diagram view, the machine is depicted as returned by \texttt{sm-graph}. The edge for the last rule used is highlighted in blue. \Cref{viztool2} depicts the same machine state as in \Cref{viztool1} in state diagram view. In both views, the right column allows for machine editing, the left column is used to provide an input word and step through machine execution, the top displays the input word, and the bottom displays the transition rules. For further details on machine execution visualization in \fsm{}, the reader is referred to a previous publication \cite{fsm-viz}.

\subsection{An Illustrative Example}
\label{init-ex}

To illustrate programming in \fsm{}, consider the following example:
\begin{alltt}
     ;; L(LNDFA) = {\(\epsilon\)} \(\cup\) aa\(\sp{\texttt{*}}\) \(\cup\) ab\(\sp{\texttt{*}}\)
     (define LNDFA (make-ndfa \quot{}(S A B F)
                              \quot{}(a b)
                              \quot{}S
                              \quot{}(A B F)
                              \qquot{}((S a A)
                                (S a B)
                                (S ,EMP F)
                                (A b A)
                                (B a B))))

     ;; Tests for LNDFA
     (check-equal? (sm-apply LNDFA \quot{}(a b a)) \quot{}reject)
     (check-equal? (sm-apply LNDFA \quot{}(b b b b b)) \quot{}reject)
     (check-equal? (sm-apply LNDFA \quot{}(a b b b b a a a)) \quot{}reject)
     (check-equal? (sm-apply LNDFA \elist) \quot{}accept)
     (check-equal? (sm-apply LNDFA \quot{}(a)) \quot{}accept)
     (check-equal? (sm-apply LNDFA \quot{}(a a a a)) \quot{}accept)
     (check-equal? (sm-apply LNDFA \quot{}(a b b)) \quot{}accept)
\end{alltt}
The language of this machine contains the empty word and all words that start with an \texttt{a} and end with either an arbitrary number of \texttt{a}s or an arbitrary number of \texttt{b}s. Essentially, the machine nondeterministically decides if the given word is empty, in \texttt{ab$^{\texttt{*}}$}, in \texttt{aa$^{\texttt{*}}$}, or is rejected. The tests validate \texttt{LNDFA}.

Machine traces are:
\begin{alltt}
     > (sm-showtransitions LNDFA \elist{})
     \quot{}((() S) (() F) accept)
     > (sm-showtransitions LNDFA \quot{}(a a a a))
     \quot{}(((a a a a) S) ((a a a) B) ((a a) B) ((a) B) (() B) accept)
     > (sm-showtransitions LNDFA \quot{}(a b b))
     \quot{}(((a b b) S) ((b b) A) ((b) A) (() A) accept)
     > (sm-showtransitions LNDFA \quot{}(a b b a))
     \quot{}reject
\end{alltt}
The trace is a list of machine configurations ending with the result. Each configuration is represented as a list containing the unconsumed input and the machine's state. The last example does not display the configurations traversed because \texttt{LNDFA} is nondeterministic and rejects.

\section{The Transformation in \fsm{}}
\label{fsm-transformation}

\subsection{Design Idea}

For the discussion on transforming an \ndfa{} to a \dfa{}, let \texttt{N = (make-ndfa S \sig{} s F \delt)}. The transformation hinges on first computing a transition function between super states and then encoding each super state as a \dfa{} state. Once both the transition function and the encoding for super states are developed,  the \dfa{} is constructed as follows:
\begin{alltt}
 (make-dfa S\(\sp{\prime}\)=<encoding of super states>
           \sig{}
           s\(\sp{\prime}\)=<encoding of E(s)>
           F\(\sp{\prime}\)=<encoding of super states that contain f\(\in\)F>
           \delt{}\(\sp{\prime}\)=<transition function between encoded super states>)
\end{alltt}

Instead of defining \texttt{S}$^{\prime}$ as 2$^{\texttt{S}}$, super state transitions are first computed and then from it the \dfa{}'s super states are extracted. To compute the transition function, each known super state, \texttt{P}, must be processed. At the beginning, the only known super state is \texttt{E(s)}. For each state, \texttt{p$\in$P}, each alphabet element, \texttt{a}, is processed to compute possibly new super states. The union of the super states obtained from processing \texttt{a} for each \texttt{p$\in$P} is the super state the \dfa{} moves to from \texttt{P} on an \texttt{a}. For instance, let \texttt{P = \{p$_1$ p$_2$ p$_3$\}} and let \texttt{(p$_1$ a r),(p$_1$ a s),(p$_3$ a t)$\in$\delt}. On an \texttt{a}, \texttt{N} may transition from \texttt{p$_1$} to any state in \texttt{E(r)$\cup$E(s)}, from \texttt{p$_2$} nowhere, and from \texttt{p$_3$} to any state in \texttt{E(t)}. Therefore, we may describe a transition in the \dfa{} as follows:
\begin{alltt}
     ((p\(\sb{1}\) p\(\sb{2}\) p\(\sb{3}\)) a (E(r) \(\cup\) E(s)\(\cup\) E(t)))
\end{alltt}
That is, the \dfa{} transitions from super state \texttt{P} on an \texttt{a} to a super state \texttt{Q}, where \texttt{Q = (E(r) $\cup$ E(s) $\cup$ E(t))}. Once the transition function between super states has been computed, it is a straightforward matter to extract the super states and generate \fsm{} symbols to encode them. This encoding is then used to create the states, the starting state, the final states, and the rules for use with \texttt{make-dfa}.

\subsection{Illustrative Example}

\begin{figure}[t!]
\centering
\begin{subfigure}{0.3\textwidth}
\centering
\begin{alltt}
(define ND
  (make-ndfa
    \quot{}(S A B C D E)
    \quot{}(a b)
    \quot{}S
    \quot{}(S)
    \qquot{}((S a A)
      (S a B)
      (A b C)
      (B b D)
      (C a E)
      (D ,EMP S)
      (E ,EMP S))))
\end{alltt}
\caption{Implementation.}
\label{samplendfa-imp}
\end{subfigure}
\hfill
\begin{subfigure}{0.6\textwidth}
\centering
\includegraphics[scale=0.50]{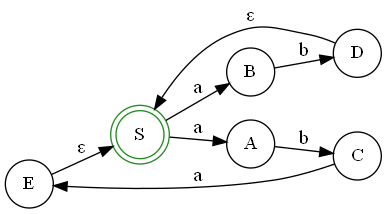}
\caption{Transition diagram.}
\label{samplendfa-graph}
\end{subfigure}
\caption{An \ndfa \ for \texttt{L = (aba $\cup$ ab)$^{\texttt{*}}$.}} \label{sample-ndfa}
\end{figure}

To illustrate the proposed constructor, consider the \ndfa{} displayed in \Cref{sample-ndfa}. First, the empties for each state are computed. The following table displays the results:
\begin{center}
\begin{tabular}{|c|l|}
  \hline
  state & E(state) \\ \hline
  S & (S) \\ \hline
  A & (A) \\ \hline
  B & (B) \\ \hline
  C & (C) \\ \hline
  D & (D S) \\ \hline
  E & (E S) \\ \hline
\end{tabular}
\end{center}

Second, the super state transition function is computed. The process starts with the only known super state, \texttt{E(S)}, which is the start super state for the \dfa. For each element, \texttt{a}, of the alphabet, the union of the states that can be reached from any states in \texttt{E(S)} by first consuming \texttt{a} is computed. \texttt{E(S)} only contains \texttt{S} and, thus, the following transitions are obtained:
\begin{alltt}
     ((S) a (A B))
     ((S) b ())
\end{alltt}
Observe that two new needed super states have been discovered: \texttt{(A B)} and \elist{} (denoting the dead state for the \dfa{} under construction). These are added to the list of unprocessed super states. The process is repeated for each unprocessed super state. For instance, the process may continue with \texttt{(A B)}. Both \texttt{A} and \texttt{B} can go nowhere on a \texttt{a}. On a \texttt{b}, from \texttt{A} the machine can transition to \texttt{C} and from \texttt{B} the machine can transition to \texttt{D} and by an \et{} to \texttt{S}. The needed super state transitions are:
\begin{alltt}
     ((A B) a ())
     ((A B) b (C D S))
\end{alltt}
In this case, a single new super state is discovered, \texttt{(C D S)}, and added to the list of unprocessed super states. The process continues in a similar fashion until there are no more super states to process. The following table summarizes the computed super state transition function:
\begin{center}
\begin{tabular}{|l|l|l|}
  \hline
  \texttt{Super State} & \texttt{a}         & \texttt{b} \\ \hline
  (S)         & (A B)     & () \\ \hline
  (A B)       & ()        & (C D S) \\ \hline
  (C D S)     & (E S A B) & () \\ \hline
  (E S A B)   & (A B)     & (C D S) \\ \hline
  ()          & ()        & () \\ \hline
\end{tabular}
\end{center}

\begin{figure}[t!]
\centering
\begin{subfigure}{0.4\textwidth}
\centering
\begin{alltt}
(define D
  (make-dfa \qquot(S A B C ,DEAD)
            \quot(a b)
            \quot{}S
            \quot(S B C)
            \qquot((S a A) (S b ,DEAD)
              (A a ,DEAD) (A b B)
              (B a C) (B b ,DEAD)
             (C a A) (C b B)
             (,DEAD a ,DEAD)
             (,DEAD b ,DEAD))))

;; Tests for D
(check-equal?
  (sm-testequiv? D ND 500)
  #t)
(check-equal?
  (sm-testequiv? (ndfa->dfa ND)
                 D
                 500)
  #t)
\end{alltt}
\caption{Implementation.}
\label{sampledfa-imp}
\end{subfigure}
\hfill
\begin{subfigure}{0.5\textwidth}
\centering
\includegraphics[scale=0.45]{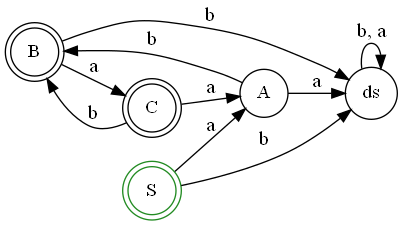}
\caption{Transition diagram.}
\label{samplen=dfa-graph}
\end{subfigure}
\caption{A \dfa{} for \texttt{L = (aba $\cup$ ab)$^{\texttt{*}}$}.} \label{sample-dfa}
\end{figure}

To build a \dfa{} the super states must be mapped to \fsm{} states. The following table is one such encoding
\footnote{\texttt{DEAD} is an \fsm{} variable denoting the default dead state name.}:
\begin{center}
\begin{tabular}{|l|c|}
  \hline
  \texttt{Super State} & \dfa \ \texttt{State}  \\ \hline
  (S)         & S         \\ \hline
  (A B)       & A         \\ \hline
  (C D S)     & B         \\ \hline
  (E S A B)   & C         \\ \hline
  ()          & DEAD      \\ \hline
\end{tabular}
\end{center}
Observe that \texttt{S}, \texttt{B}, and \texttt{C} are final states in the \dfa{} because the super state they represent contains the only final state in \texttt{ND}. This leads to the \dfa{} displayed in \Cref{sample-dfa}. The tests validate that the deterministic and nondeterministic implementations decide the same language. The second \texttt{check-equal?} uses \fsm's primitive, \texttt{ndfa->dfa}, to transform an \ndfa{} to a \dfa{} and validates that \fsm's transformation and \texttt{D} are equivalent.

\subsection{Implementation Sketch}

The implementation presented uses the following data definitions:
\begin{alltt}
     ;; Data Definitions
     ;;
     ;;  An ndfa transition rule, ndfa-rule, is a
     ;;  (list state symbol state)
     ;;
     ;;  A super state, ss, is a (listof state)
     ;;
     ;;  A super state \dfa rule, ss-dfa-rule, is a
     ;;  (list ss symbol ss)
     ;;
     ;;  An empties table, emps-tbl, is a
     ;;  (listof (list state ss))
     ;;
     ;;  A super state name table, ss-name-table, is a
     ;;  (listof (list ss state))
\end{alltt}

\begin{figure}[t!]
\begin{alltt}
;; ndfa \arrow dfa
;; Purpose: Convert the given ndfa to an equivalent dfa
(define (ndfa2dfa M)
 (if (eq? (sm-type M) \quot{}dfa)
   M
   (convert (sm-states M) (sm-sigma M) (sm-start M) (sm-finals M) (sm-rules M))))

;; Tests for ndfa2dfa
(define M (ndfa2dfa AT-LEAST-ONE-MISSING))
(check-equal? (sm-testequiv? AT-LEAST-ONE-MISSING M 500) #t)
(check-equal? (sm-testequiv? M (ndfa->dfa AT-LEAST-ONE-MISSING) 500) #t)

(define N (ndfa2dfa ND))
(check-equal? (sm-testequiv? ND N 500) #t)
(check-equal? (sm-testequiv? N (ndfa->dfa ND) 500) #t)
\end{alltt}
\caption{The main function to convert an \ndfa{} to a \dfa.}\label{ndfa2dfa-def}
\end{figure}

The main function, \texttt{ndfa2dfa}, tests if the given machine is a \dfa{}. If so, it returns it given that no transformation is needed. Otherwise, an auxiliary converting function is called with the components of the given \ndfa{}. The implementation is displayed in \Cref{ndfa2dfa-def}.

\begin{figure}[t!]
\begin{alltt}
;; (listof state) alphabet state (listof state) (listof ndfa-rule) \arrow dfa
;; Purpose: Create a dfa from the given ndfa components
(define (convert states sigma start finals rules)
  (let* [(empties (compute-empties-tbl states rules))
         (ss-dfa-rules
          (compute-ss-dfa-rules (list (extract-empties start empties))
                                sigma
                                empties
                                rules
                                \elist))
         (super-states (remove-duplicates
                         (append-map (\lamb (r) (list (first r) (third r)))
                                     ss-dfa-rules)))
         (ss-name-tbl (compute-ss-name-tbl super-states))]
    (make-dfa (map (\lamb (ss) (second (assoc ss ss-name-tbl)))
                   super-states)
              sigma
              (second (assoc (first super-states) ss-name-tbl))
              (map (\lamb (ss) (second (assoc ss ss-name-tbl)))
                   (filter (\lamb (ss) (ormap (\lamb (s) (member s finals)) ss))
                           super-states))
              (map (\lamb (r) (list (second (assoc (first r) ss-name-tbl))
                                    (second r)
                                    (second (assoc (third r) ss-name-tbl))))
                   ss-dfa-rules)
              \quot{}no-dead)))
\end{alltt}
\caption{The \texttt{convert} implementation.} \label{convert}
\end{figure}

The \texttt{convert} function takes as input the components of an \ndfa{}. To build a \dfa{} it computes the following values:
\begin{enumerate}
  \item The empties table
  \item The super state transition rules
  \item The super state name table
\end{enumerate}
The implementation of this design is displayed in \Cref{convert}. The functions \texttt{compute-empties-tbl} and \texttt{compute-ss-name-tbl} are used to compute, respectively, the empties table and the super states table. These functions are straightforward to implement and, in the interest of brevity, are not presented.

\begin{figure}[t!]
\begin{alltt}
;; (listof ss) alphabet emps-tbl (listof ndfa-rule) (listof ss)
;; \arrow (listof ss-dfa-rule)
;; Purpose: Compute the super state dfa rules
;; Accumulator Invariants:
;;               ssts = the super states explored
;;     to-search-ssts = the super states that must still be explored
(define (compute-ss-dfa-rules to-search-ssts sigma empties rules ssts)
      \vdotss
 (if (empty? to-search-ssts)
     \elist
     (let* [(curr-ss (first to-search-ssts))
            (reachables (find-reachables curr-ss sigma rules empties))
            (to-super-states
             (build-list (length sigma) (\lamb (i) (get-reachable i reachables))))
            (new-rules (map (\lamb (sst a) (list curr-ss a sst))
                            to-super-states
                            sigma))]
       (append
         new-rules
         (compute-ss-dfa-rules
           (append (rest to-search-ssts)
                   (filter (\lamb (ss)
                             (not (member ss (append to-search-ssts ssts))))
                           to-super-states))
           sigma
           empties
           rules
           (cons curr-ss ssts))))))
\end{alltt}
\caption{Function to compute the super state transition function.} \label{sstf}
\end{figure}

To compute the super state transition function we perform a breadth-first search rooted at the super states that still need to be explored to identify transition rules. There are two accumulators: one for the super states left to explore and one for the super states already explored. If there are no states left to explore, the empty list is returned because the search for super state \dfa{} rules is done. Otherwise, the first unexplored super state, denoted \texttt{curr-ss}, is processed. The set of reachable states for every state in the first unexplored super state, consuming every element of the alphabet, is computed. For instance, if the first unexplored super state is \texttt{\quot{}(A B C)} and the alphabet is \texttt{\quot{}(a b)} then the reachable states have the following structure:
\begin{alltt}
     (((reachable from A on a) (reachable from A on b))
      ((reachable from B on a) (reachable from B on b))
      ((reachable from C on a) (reachable from C on b)))
\end{alltt}
There are three sublists in the list: one for each state in the first unexplored super state. Each sublist has a list of states that are reachable for each element of the alphabet. From this, the super states to transition into, \texttt{to-super-states}, are computed. For instance, the super state reachable on an \texttt{a} is formed by all the states reachable on an \texttt{a} from \texttt{A}, \texttt{B}, and \texttt{C} (without repetitions). The \dfa{} transition rules are generated by simultaneously traversing \texttt{to-super-states} and the alphabet. Assuming \texttt{a} is the current alphabet element and \texttt{sst} is the current element of \texttt{to-super-states} during such a traversal, then a super state \dfa{} rule of the following form is created: \texttt{(curr-ss a sst)}. Finally, the super state \dfa{} rules generated are appended with the result of recursively processing a new accumulator for unexplored super states containing the rest of the said accumulator and any new super states generated for the new rules, the same alphabet, the same empties table, the same list of \ndfa{} rules, and a new accumulator for explored states obtained by adding the first unexplored super state. The \texttt{compute-ss-dfa-rules} function based on this design is displayed in \Cref{sstf}. This is the bulk of what is needed to compute the super state \dfa \ transition function. The remaining auxiliary functions are straightforward and not further discussed.

\section{Visualization in \fsm{}}
\label{fsm-viz}

To visualize the \ndfa{} to \dfa{} transformation, two state diagrams are built at every step. The first is for the \ndfa{} and is used to highlight the edges that have been used in the construction of the \dfa{}. The second is for the \dfa{} that is built piecemeal one edge at a time.

\subsection{Design Idea}

The \ndfa{} is denoted by \texttt{N} and the \dfa{} is denoted by \texttt{D}. The transition diagram for \texttt{N} is always fully displayed. The transition diagram for \texttt{D} is built one transition at a time. The user may step forwards or backwards in the computation one step at the time. In addition, the user may have the \dfa{}'s transition diagram completed in one step.

The \ndfa{}'s edges are partitioned into three subsets that are color coded. The first is the set of edges that have not yet been used in the \dfa{}'s construction. These edges are drawn in black and are referred to as \texttt{bledges}. The second is the set of edges used to add the latest \dfa{} edge. These edges are highlighted in violet and are referred to as \texttt{hedges}. Finally, the third set is for edges that have previously been used in the \dfa{} construction. These edges are faded out in gray and are referred to as \texttt{fedges}.

The state transition diagram is built using the output of \texttt{compute-ss-dfa-rules}. The \dfa{} super state transitions are partitioned into two sets. The first contains the super state transitions that have been processed in the construction and are denoted as \texttt{ad-edges}. The second contains the super state transitions that are unprocessed and are denoted by \texttt{up-edges}. In addition, the super states included in the constructed part of the \dfa{} are accumulated to prevent adding any given super state more than once.

Initially, \texttt{D}'s transition diagram only contains \texttt{E(s)}, where \texttt{s} is \texttt{N}'s starting state. In \texttt{N}'s transition diagram, initially, the \ets{} on any path from \texttt{s} to any state in \texttt{E(s)} are \texttt{hedges} and all other edges are \texttt{bledges}. Every time the user moves the computation forward, the next element of \texttt{up-edges}, \texttt{(SS$_1$ a SS$_2$)}, is processed and added to \texttt{ad-edges} resulting in a new edge appearing in the \dfa{}'s transition diagram. Simultaneously, the \texttt{hedges} are added to the \texttt{fedges} to have them faded. In addition, the new set of \texttt{hedges} and the new set of \texttt{bledges} are computed. 
The new set of \texttt{hedges} contains all edges in \texttt{N} that consume \texttt{a} from any state in \texttt{SS$_1$} and all the \ets{} that are reachable from any state in \texttt{SS$_1$} after consuming the \texttt{a}. The new set of \texttt{bledges} is obtained by removing the new \texttt{hedges} from it. Finally, \texttt{SS$_2$} is added, if not already present, to the accumulator for super states that are part of the \dfa{} constructed thus far.

Every time the user moves the computation backwards, the process above is reversed. The last edge added is removed from \texttt{D}'s transition diagram. That is, the newest element in \texttt{ad-edges} is moved to \texttt{up-edges}. The next newest element in \texttt{ad-edges} is used to compute the new sets of \texttt{hedges}, \texttt{fedges}, and \texttt{bledges} as described above. If necessary, the newest element in the super state accumulator is removed.

Finally, when the user moves to complete the \ndfa{} transformation, all the \texttt{up-edges} are processed as described above. In this manner, the user may still step backwards in the computation if she desires.

\subsection{Illustrative Example}

\begin{figure}[t!]
\centering
\begin{subfigure}{0.3\textwidth}
\centering
\begin{alltt}
(define aa-ab
  (make-ndfa
    \quot{}(S A B F)
    \quot{}(a b)
    \quot{}S
    \quot{}(A B)
    \qquot{}((S a A)
      (S a B)
      (S ,EMP F)
      (A a A)
      (B b B))))
\end{alltt}
\caption{Implementation.}
\label{aaab-imp}
\end{subfigure}
\hfill
\begin{subfigure}{0.6\textwidth}
\centering
\includegraphics[scale=0.50]{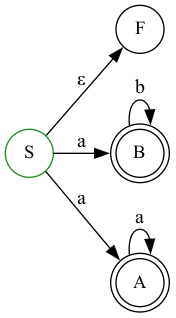}
\caption{Transition diagram.}
\label{aaab-graph}
\end{subfigure}
\caption{An \ndfa{} for \texttt{L = L=\{$\epsilon \cup aa^{\texttt{*}} \cup ab^{\texttt{*}}$\}}} \label{aa-ab-ndfa-graph}
\end{figure}

\begin{figure}[t!]
\centering
\includegraphics[scale=0.3]{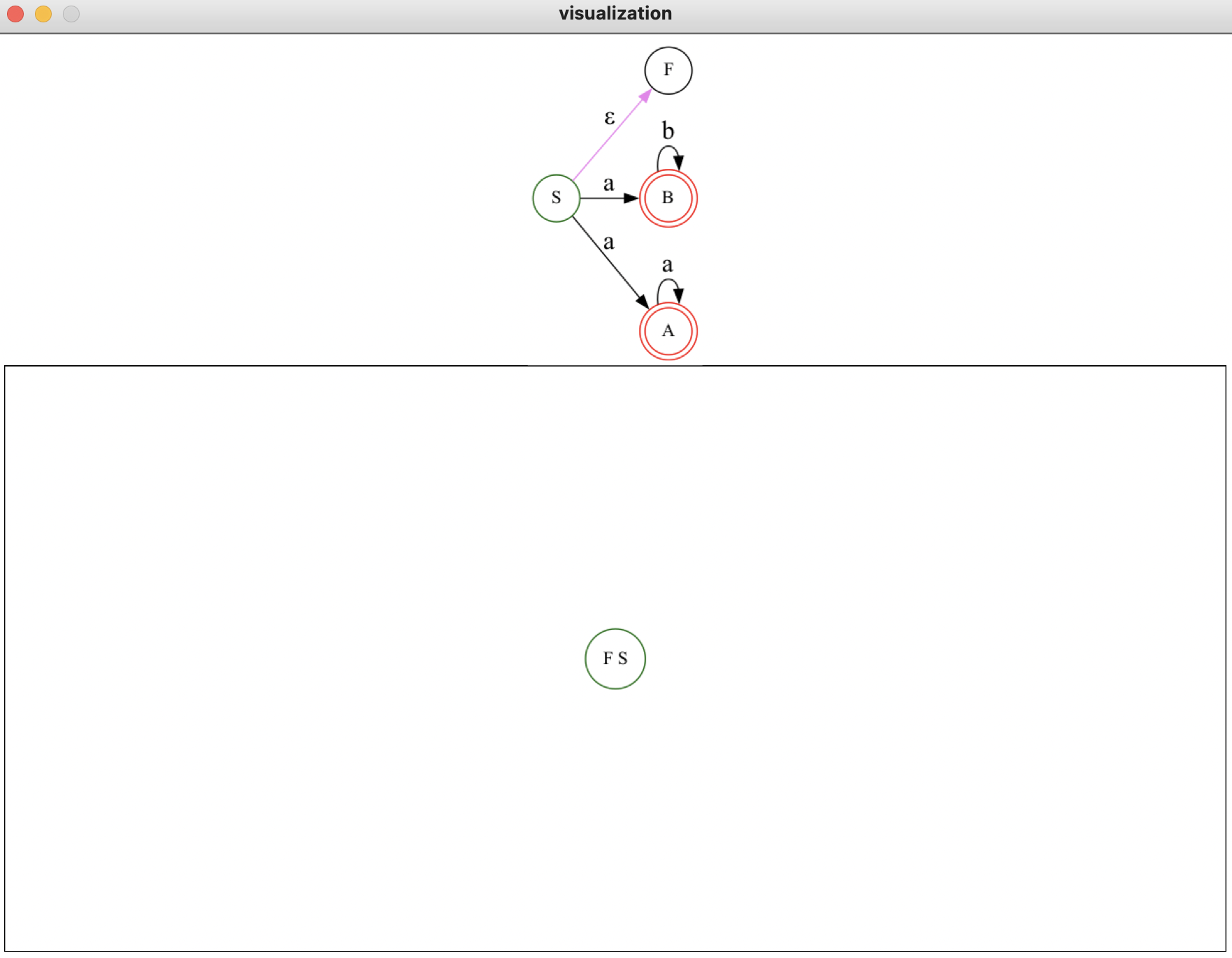}
\caption{The initial visualization state.} \label{initial-viz}
\end{figure}

To illustrate the transformation, consider the \ndfa{} defined and its transition diagram in \Cref{aa-ab-ndfa-graph}. Initially, the diagram for the \dfa{} only contains one super state for \texttt{E(S)=(F S)}. The diagram for the \ndfa{} only has one \texttt{hedge} from \texttt{S} to \texttt{F}, because \texttt{(S $\ep$ F)} is the only edge used so far in the \dfa{}'s construction. All other edges in the ndfa{}'s diagram are \texttt{bledges}. This initial state of the visualization is displayed in \Cref{initial-viz}.

\begin{figure}[t!]
\centering
\includegraphics[scale=0.3]{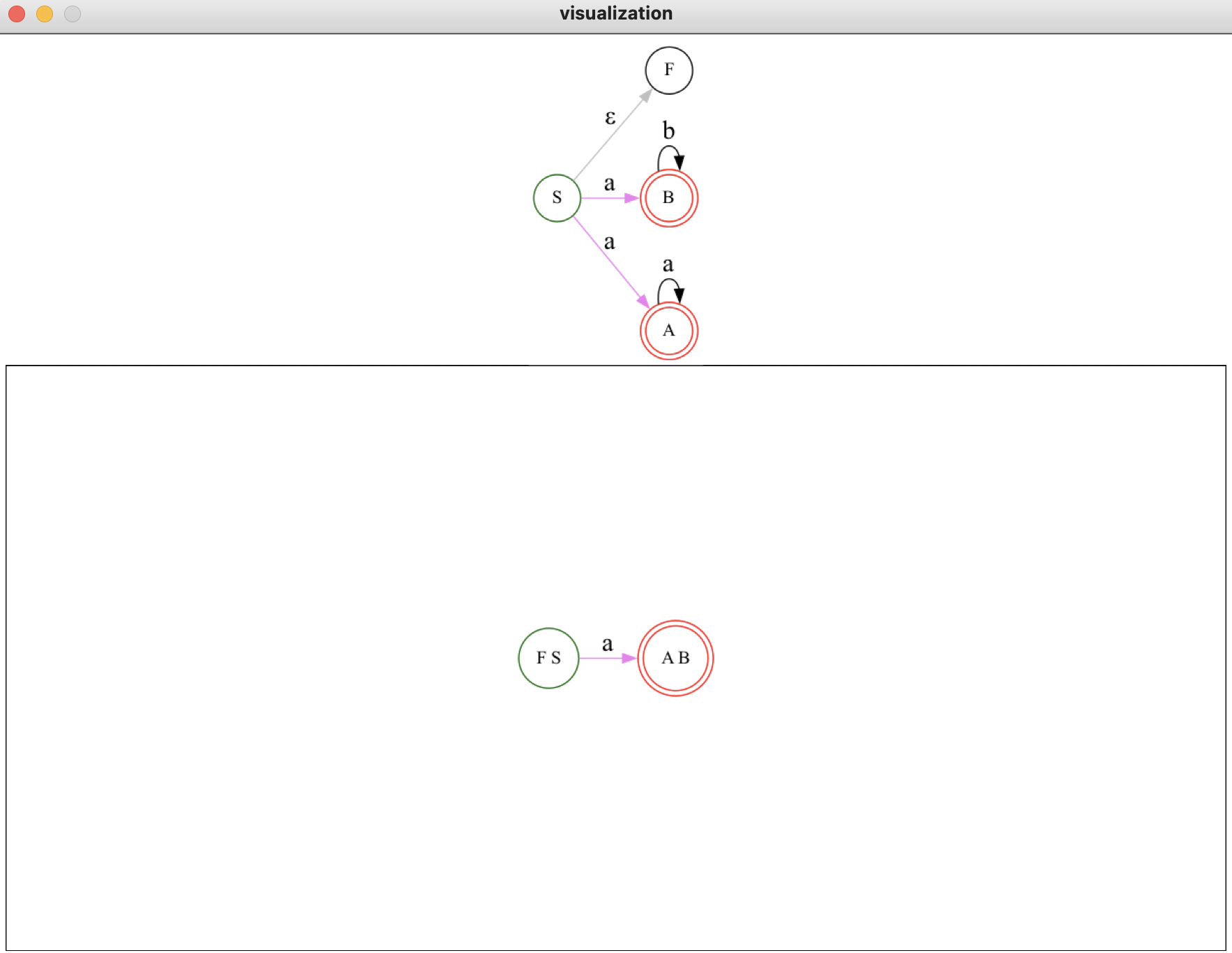}
\caption{The visualization state after moving the visualization one step forward.} \label{moving-forward}
\end{figure}

When the user moves one step forward in the computation, the simulation adds \texttt{((F S) a (A B))} to the \dfa{}. It is highlighted in violet as it is the last \dfa{} edge added. In the \ndfa{}'s diagram, the $\epsilon$-transition from \texttt{S} to \texttt{F} becomes a \texttt{fedge} and, thus, is highlighted in gray. In addition, all reachable edges out of \texttt{S} on an \texttt{a} are part of the \dfa{}'s transition and, thus, are \texttt{hedge}s highlighted in violet. All other \ndfa{} edges are \texttt{bledge}s rendered in black. Observe that the user can easily determine why there is a \dfa{} super state \texttt{(A B)}. Its existence follows from the \texttt{hedge}s highlighted in violet in the \ndfa{}. These \texttt{hedge}s  define all the states reachable from \texttt{S} on an \texttt{a}.

\begin{figure}[t!]
\centering
\includegraphics[scale=0.3]{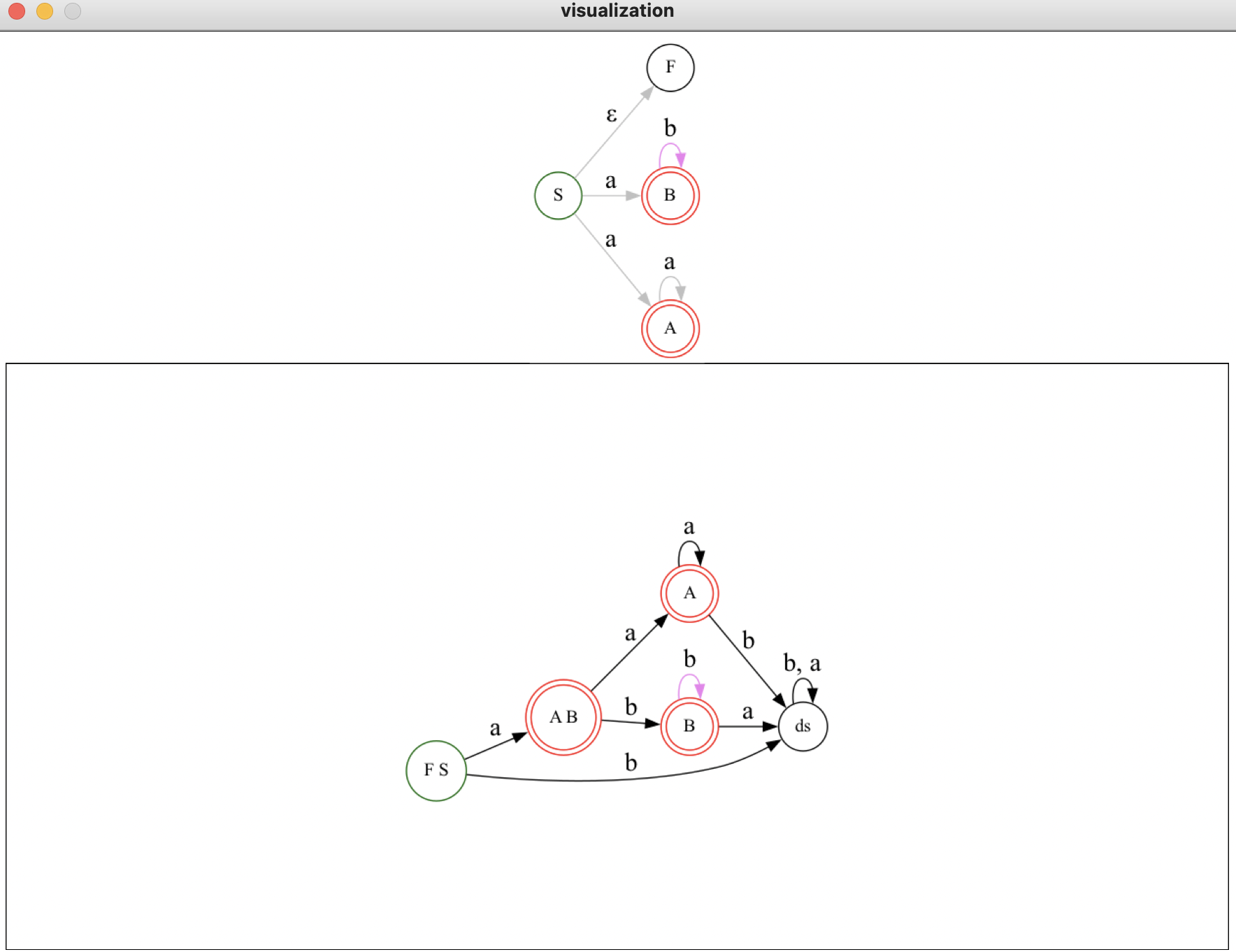}
\caption{The final visualization state.} \label{complete-computation}
\end{figure}

If the user completes the computation in one step, all the \dfa{} edges are rendered. The \dfa{} transition diagram is generated by processing all the super state transitions in the same order as if the user stepped through the computation. Therefore, the \dfa{} edge for the last super state transition processed is highlighted in violet. This is depicted in the finalized \dfa{} transition diagram in \Cref{complete-computation}. In the diagram for the \ndfa{}, only \texttt{(B b B)} is a \texttt{hedge} given that the only state reachable from \texttt{B} on a \texttt{b} is \texttt{B}. Observe that this clearly explains why there is a \dfa{} super state that only contains \texttt{B}. The reader can compare the \dfa{} transition diagram in \Cref{complete-computation} with the transition diagrams in \Cref{jflap2} and \Cref{openflap2} obtained using, respectively, \texttt{JFLAP} and \texttt{OpenFLAP}. \fsm{}'s use of \gviz{} results in superior renderings for \dfa{} transition diagrams.

\begin{figure}[t!]
\centering
\includegraphics[scale=0.3]{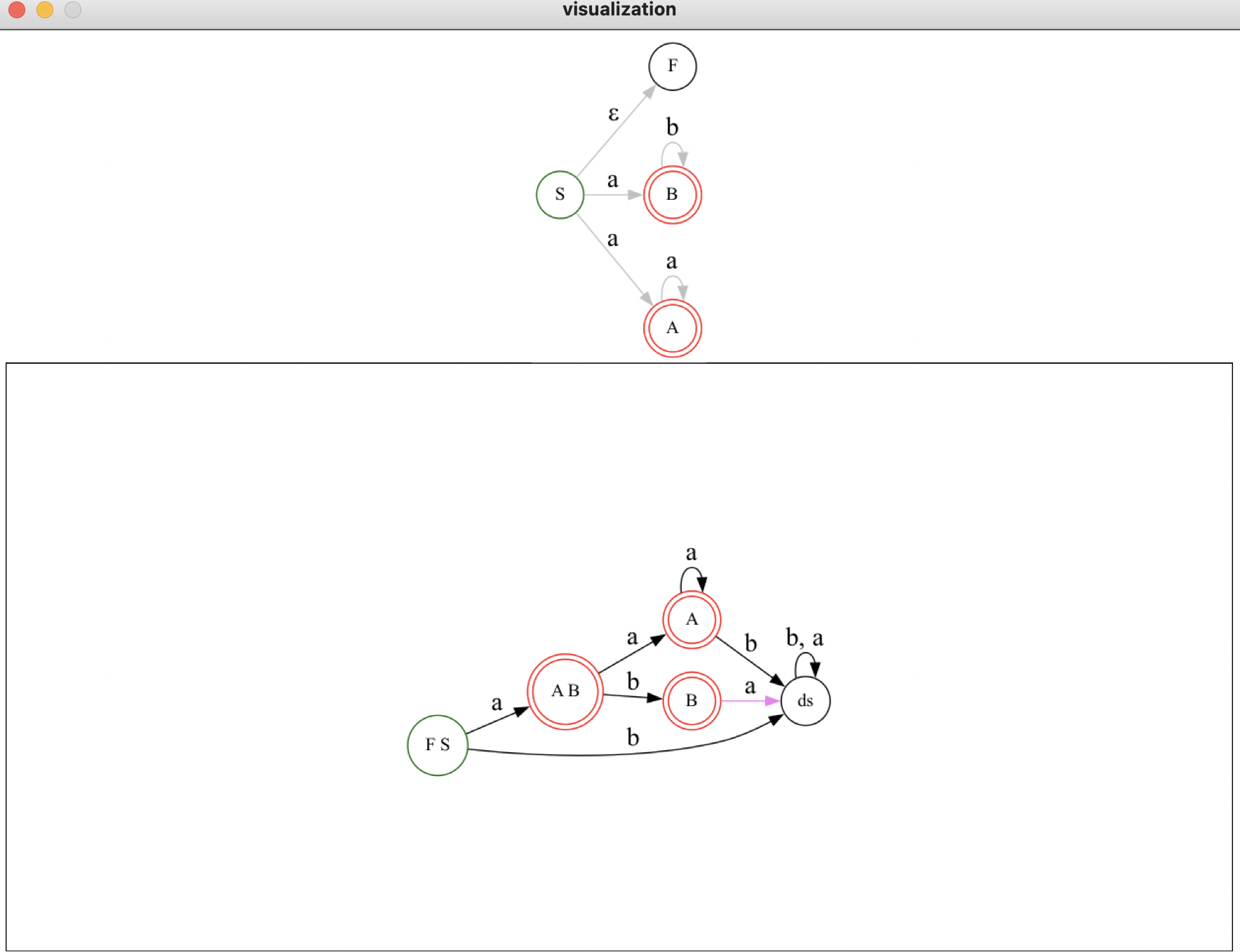}
\caption{The visualization state after moving the computation one step backwards.} \label{move-backwards}
\end{figure}

Finally, we illustrate what happens when the user takes a step backwards in the computation. The last edge added to the \dfa{}'s diagram is removed, making the next-to-last \dfa{} edge added the last edge added in the new state of the computation. For the \ndfa{}, the current \texttt{hedge}s are made \texttt{bledge}s and, thus, are no longer highlighted. In addition, the \texttt{hedge}s that existed when the next-to-last \dfa{} edge  was added are recomputed and removed from \texttt{fedge}s. For instance, if the user moves one step backwards in the computation's state displayed in \Cref{complete-computation}, the \dfa{} edge \texttt{(B b B)} is removed and the next-to-last \dfa{} edge added, \texttt{(B a ds)}, is highlighted. In the \ndfa{} \texttt{(B b B)} becomes a \texttt{fedge}. This is correct because this edge is part of the \dfa{} construction given the super state edge \texttt{((A B) b B)}. This illustrates that multiple \ndfa{} transitions end up in the same super state, thus, removing an edge from the \dfa{} diagram does not automatically remove an edge from the \ndfa{} diagram. Sometimes an \ndfa{} edge is moved to a different subset of the \ndfa{}-edges partition. To close, we note that when the last \dfa{} edge added is to the fresh dead state then there are no \texttt{hedge}s in the \ndfa{} given that this state is not part of the \ndfa{}.


\subsection{Implementation Highlights}

The state of the visualization is represented using a structure:
\begin{alltt}
   (struct viz-state (up-edges ad-edges incl-nodes M hedges fedges bledges))
\end{alltt}
The first three fields are for the state of the \dfa{} under construction described as follows:
\begin{description}
  \item[\texttt{up-edges}:] the unprocessed \dfa{} edges
  \item[\texttt{ad-edges}:] the processed \dfa{} edges
  \item[\texttt{incl-nodes}:] the super states appearing in \texttt{ad-edges}
\end{description}
The fourth field is the given \ndfa{}. The last three fields represent the partition of \ndfa{} edges described as follows:
\begin{description}
  \item[hedges:] the most recent \ndfa{} edges processed that are rendered in violet
  \item[fedges:] the previous \ndfa{} edges processed that are rendered in gray
  \item[bledges:] the unprocessed \ndfa{} that are rendered in black
\end{description}

At the beginning of the visualization, all the \dfa{} edges are unprocessed. The only super state rendered for the \dfa{} is the starting super state. For the \ndfa{}, the set of \texttt{hedge}s contains all \ndfa{} edges reachable on \ets{} from any state in the starting super state, the set of \texttt{fedge}s is empty (i.e., there are no previously processed \ndfa{} edges), and the set of \texttt{bledge}s are all the \ndfa{} edges that are not a \texttt{hedge}s. Thus, the initial \texttt{viz-state} value is defined as follows:
\begin{alltt}
  (let* [(ss-edges (compute-ss-edges M))
         (super-start-state
           (compute-empties (list (sm-start M)) (sm-rules M) \elist{}))
         (init-hedges (compute-all-hedges (sm-rules M) super-start-state \elist{}))
    (viz-state ss-edges
               \elist{}
               (list super-start-state)
               M
               init-hedges
               \elist{}
               (remove init-hedges (sm-rules M)))
\end{alltt}
Here, \texttt{compute-ss-edges} returns the super state \dfa{} edges given an \ndfa{}, \texttt{compute-empties} returns the empties of the given state, and \texttt{compute-all-hedges} returns the set of \texttt{hedge}s for the given super state and the given \dfa{} super state edge added. In this case, \elist{} is given as the edge added because there is no \dfa{} edge being added.

Moving the computation forward means that the next super state \dfa{} edge is processed. Therefore, the current \dfa{} super state edge (i.e., the first unprocessed) is moved to the set of processed \dfa{} super state edges. Any super states in the first unprocessed \dfa{} edge, if not already included, are added to \texttt{incl-nodes}. For the \ndfa{}, the new set of \texttt{hedge}s is computed using the current \dfa{} rule's destination super state. To compute the new set of \texttt{fedge}s, two cases are distinguished. When the new set of unprocessed \dfa{} super state edges is empty, the computation can no longer move forward. In order for the simulation to properly move backwards, the new \texttt{hedge}s are added to the \texttt{fedge}s. This is consistent with the design, because for edge rendering \texttt{hedge}s take precedence over \texttt{fedge}s. Thus, the rendering of the \dfa{} remains correct. When the new set of unprocessed \dfa{} super state edges is not empty, the existing \texttt{hedge}s are added to the set of \texttt{fedge}s. Finally, the new set of \texttt{bledge}s is obtained by removing the new \texttt{hedge}s. This design is implemented as follows:
\begin{alltt}
(let* [(curr-dfa-ss-edge (first (viz-state-up-edges a-viz-state)))
       (new-up-edges (rest (viz-state-up-edges a-viz-state)))
       (new-ad-edges (cons curr-dfa-ss-edge (viz-state-ad-edges a-viz-state)))
       (new-incl-nodes
         (add-included-node (viz-state-incl-nodes a-viz-state)
         curr-dfa-ss-edge))
       (new-hedges (compute-all-hedges (sm-rules (viz-state-M a-viz-state))
                                       (third curr-dfa-ss-edge)
                                       curr-dfa-ss-edge))
       (new-fedges (if (empty? new-up-edges)
                       (append new-hedges (viz-state-fedges a-viz-state))
                       (append (viz-state-hedges a-viz-state)
                               (viz-state-fedges a-viz-state))))
       (new-bledges (remove-edges new-hedges (viz-state-bledges a-viz-state)))]
  (viz-state new-up-edges
             new-ad-edges
             new-incl-nodes
             (viz-state-M a-viz-state)
             new-hedges
             new-fedges
             new-bledges))
\end{alltt}

Moving the computation backwards means that the last processed \dfa{} super state edge is moved to the set of unprocessed \dfa{} super state edges. If necessary, the last processed \dfa{} edge's destination state is removed from \texttt{incl-nodes}. For the new \texttt{hedge}s set, two cases are distinguished. When the set of processed \dfa{} super state edges is empty, the visualization is at the beginning. In this case, the new set of \texttt{hedge}s is computed using the \dfa{}'s starting super state. Otherwise, the new set of \texttt{hedge}s is computed using the first \dfa{} super state edge in the new set of processed edges. For the new set of \texttt{fedge}s, two cases are also distinguished. If the new set of processed edges is empty, then there are no \texttt{fedge}s. Otherwise, the new \texttt{fedge}s are obtained by removing  the \texttt{hedge}s computed using the destination state of the last \dfa{} super state edge processed. Finally, the set of new \texttt{bledge}s is given by appending the \texttt{hedge}s computed using the destination state of the last \dfa{} super state edge processed and the set of \texttt{bledge}s. This design is implemented as follows:
\begin{alltt}
(let* [(last-dfa-pedge (first (viz-state-ad-edges a-viz-state)))
       (new-up-edges (cons last-dfa-pedge (viz-state-up-edges a-viz-state)))
       (new-ad-edges (rest (viz-state-ad-edges a-viz-state)))
       (new-incl-nodes (remove-included-node (viz-state-incl-nodes a-viz-state)
                                             last-dfa-pedge
                                             new-ad-edges))
       (new-hedges (if (empty? new-ad-edges)
                       (compute-all-hedges
                          (sm-rules (viz-state-M a-viz-state))
                          (compute-empties
                            (list (sm-start (viz-state-M a-viz-state)))
                            (sm-rules (viz-state-M a-viz-state))
                            \elist{})
                          \elist{})
                       (compute-all-hedges (sm-rules (viz-state-M a-viz-state))
                                           (third (first new-ad-edges))
                                           (first new-ad-edges))))
       (prev-hedges (compute-all-hedges (sm-rules (viz-state-M a-viz-state))
                                            (third last-dfa-pedge)
                                            last-dfa-pedge))
       (new-fedges (if (empty? new-ad-edges)
                       empty
                       (remove-edges prev-hedges
                                     (viz-state-fedges a-viz-state))))
       (new-bledges (remove-duplicates
                      (append prev-hedges
                              (viz-state-bledges a-viz-state))))]
  (make-world new-up-edges
              new-ad-edges
              new-incl-nodes
              (viz-state-M a-viz-state)
              new-hedges
              new-fedges
              new-bledges))
\end{alltt}

Finally, when the visualization is moved to its final state all \dfa{} super state edges are processed and all super states are in \texttt{incl-nodes}. The new set of \texttt{hedge}s is computed using the destination super state of the last \dfa{} edge processed. The new set of \texttt{fedge}s is obtained by appending the \texttt{hedge}s in the same order as added by moving the visualization forward in a piecemeal fashion and the \texttt{hedge}s associated with the starting super state. Finally, the new set of \texttt{bledge}s is empty. This design is implemented as follows:
\begin{alltt}
(let* [(ss-edges (append (reverse (viz-state-ad-edges a-viz-state))
                         (viz-state-up-edges a-viz-state))))
       (super-start-state (first (first ss-edges)))
       (new-up-edges \elist{})
       (new-ad-edges (reverse ss-edges))
       (new-incl-nodes
         (remove-duplicates (append-map (\lamb{} (e) (list (first e) (third e)))
                                        new-ad-edges)))
       (new-hedges (compute-all-hedges (sm-rules (viz-state-M a-viz-state))
                                       (third (first new-ad-edges))
                                       (first new-ad-edges)))
       (new-fedges
         (append (compute-down-fedges (sm-rules (viz-state-M a-viz-state))
                 new-ad-edges)
                 (compute-all-hedges (sm-rules (viz-state-M a-viz-state))
                                     super-start-state
                                     \elist{})))
       (new-bledges \elist{})]
  (make-world new-up-edges
              new-ad-edges
              new-incl-nodes
              (viz-state-M a-viz-state)
              new-hedges
              new-fedges
              new-bledges))
\end{alltt}

\section{Concluding Remarks}
\label{concl}

This article presents a novel visualization tool for the transformation of an \ndfa{} to a \dfa{}. The visualization simultaneously renders the \ndfa{}'s transition diagram and the transition diagram for the \dfa{} constructed so far. It improves the current state-of-the-art by rendering transition diagrams in an appealing manner. In addition, the last \dfa{} edge added is highlighted as well as the corresponding \ndfa{} edges. The \ndfa{} edges that are already part of the \dfa{} construction are faded out in gray and \ndfa{} edges that are not part of the \dfa{} construction are rendered in black. In this manner, any user can easily determine why the \dfa{}'s super states exist, see how all \ndfa{} edges are processed, and comprehend the development of the \dfa{}'s transition function. The \fsm{} visualization, unlike other visualization tools for this transformation, may be advanced both forward and backwards. Finally, the \dfa{}'s transition diagram may be rendered in one step without preventing the user from moving the simulation backwards.

Future work includes adding visualization tools for other machine transformations or constructors. The targeted transformations include \dfa{} to regular expression and vice versa. The targeted constructors include the closure properties under union, concatenation, Kleene star, complement, and intersection. In addition, empirical studies will be performed to collect student feedback on the work presented in this article.

\bibliographystyle{ACM-Reference-Format}
\bibliography{ndfa2dfa-viz}


\end{document}